\documentclass[conference]{IEEEtran}
\usepackage{cite}
\usepackage[fleqn]{amsmath}
\usepackage{graphicx}
\usepackage{comment}
\usepackage{amsmath}
\usepackage{amssymb}
\usepackage{subfigure}
\usepackage{epstopdf}
\begin{document}
\title{A novel approach for droplet position sensing in electrowetting devices}
\author{\authorblockN{Shiraz Sohail\textbf{$^\ast$} and Karabi Biswas}
\authorblockA{Department of Electrical Engineering,\\ Indian Institute of Technology\\Kharagpur, India-721302.\\\textbf{$^\ast$}Email:ssohail@iitkgp.ac.in}}
\maketitle
\begin{abstract}
A droplet position sensing scheme for monitoring multiple droplets has been proposed, which gives a direct voltage output linearly proportional to droplet position in electrowetting-on-dielectric (EWOD) based devices. An extra dielectric and metal layers are required in the bottom substrate for physical realization of the scheme which provides better isolation between actuation and sensing line. The capacitors formed due to the extra dielectric and metal layers facilitate direct voltage output proportional to droplet position during transport. Energy-based model has been used for analysis purpose. Results show that the additional dielectric and metal layer in the bottom substrate does not significantly change the driving electrostatic force profile. This makes the proposed scheme compatible for digital microfluidic operations.
\end{abstract}
\IEEEpeerreviewmaketitle
\section{Introduction}
\label{one}
EWOD (electrowetting-on-dielectric) microsystems without position feedback have been developed mostly in the last decade \cite{Jean}. But, accuracy of assays in $\mu$TAS is largely determined by the droplet volume control of the reagent dosing. So, droplets are dispensed from on-chip reservoirs by sequential pinch off and it is a difficult task to achieve a synchronization between droplet position and electrode actuation. In particular, during asymmetric splitting of droplets. Sometimes, even a resistance to droplet movement may result in a complete loss of synchronization \cite{Shin, Shih, Schertzer, Murran, Bhattacharjee, Sadeghi, Luan2012}. These problems can be solved by introducing a position feedback control scheme. However, this requires integration of miniaturized position sensors with EWOD fluidic chips.
\par
Earlier works on droplet position sensing for feedback is found in \cite{Shin, Shih, Schertzer, Murran, Bhattacharjee, Sadeghi, Luan2012}. In summary, position sensing scheme has been implemented either by measuring built-in capacitance \cite{Murran, Schertzer}, introducing  integrated optical sensor (LED and photodetector) \cite{Luan2012}, using LABVIEW based machine vision mechanism \cite{Shin} or connecting a passive circuit comprising of resistors or capacitors \cite{Shih, Sadeghi, Bhattacharjee}. Capacitive sensors provide a particularly attractive option since the method of detection is non-intrusive, highly sensitive and suitable for electrically conducting or insulating liquids. However, simultaneous actuation of control electrode for droplet transport and measuring capacitance for position detection is not facilitated by most of the techniques found in literature because of interference between actuation line and sensing line \cite{Murran, Schertzer, Bhattacharjee}. Those techniques providing isolation are not compatible with multiple droplet position detection simultaneously \cite{Shih, Sadeghi} or use optical sensor \cite{Shin, Luan2012}.
\par
In this work, a scheme for monitoring multiple droplets position simultaneously in electrowetting-on-dielectric (EWOD) based devices has been proposed.  This scheme provides a direct voltage output and isolates the actuation and sensing line. Details of the scheme has been discussed in section \ref{two}. The mathematical modelling for EWOD actuation force and output voltage for droplet position has been carried out in section \ref{three}. Results have been presented and discussed in section \ref{four}, while conclusion has been provided in section \ref{five}. 
\section{Proposed Position Sensing Scheme}
\label{two}
\begin{figure}[!tbh]
\begin{center}
\includegraphics[width=0.9\columnwidth, height=2.0in]{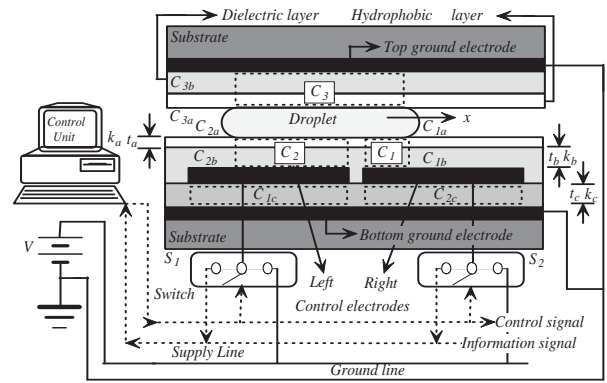}   
\end{center}
\caption{Side view of the proposed scheme for droplet position sensing.}
\label{fig1}  
\end{figure}
Side view of the proposed scheme has been shown in Fig. \ref{fig1}. This scheme consist of two metal layers, two dielectric layers and one hydrophobic layer in the bottom substrate while top substrate consist of one metal, one dielectric and one hydrophobic layer. Out of the two metal layers in the bottom substrate, the first layer is a flat thin film while the second is patterned to form coplanar electrodes for discrete droplet motion. The top and bottom flat metal layers are always connected to ground potential. The droplet is moved to right side by actuating right control electrode (Fig. \ref{fig1}), while the potential of the left control electrode is monitored to sense the droplet position and vice versa. Therefore this scheme is compatible for monitoring multiple droplet simultaneously. Note, in traditional EWOD system, first dielectric and metal layer in the bottom substrate is not found \cite{Jean}.
\begin{figure}[!tbh]
    \begin{center}
                \subfigure[]{
            \label{fig2}
            \includegraphics[width=0.5\columnwidth, height=1.7in]{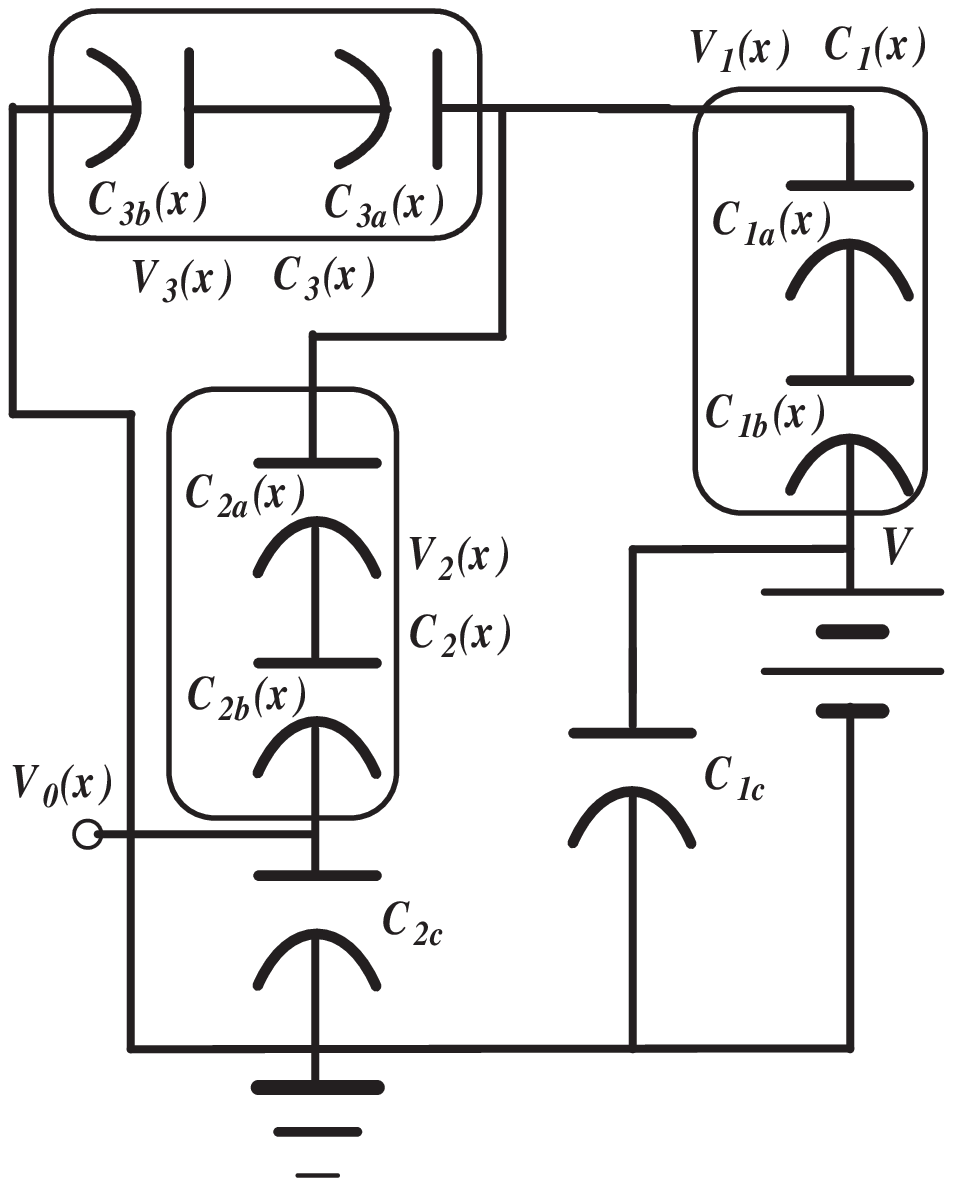}
        }
        \subfigure[]{
            \label{fig3}
            \includegraphics[width=0.4\columnwidth, height=1.7in]{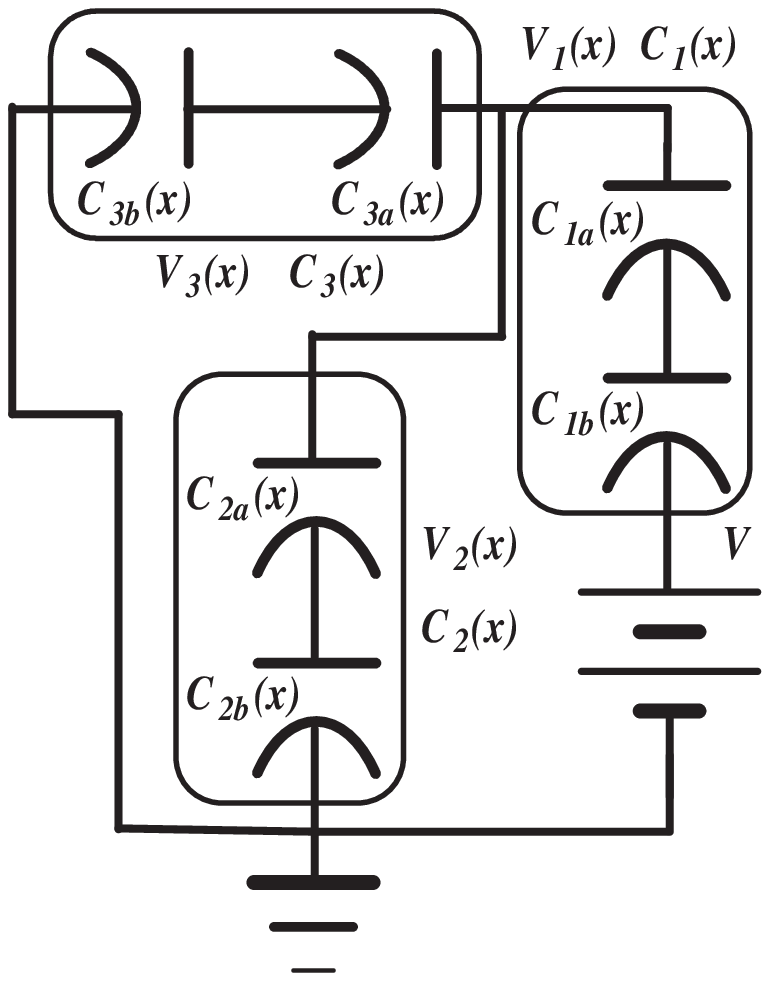}
        }
      \end{center}
\caption{Electrical equivalent circuit: (a) Proposed sensing scheme (b) Traditional EWOD scheme.}
\end{figure} 
\par
\begin{comment}
The dielectric and hydrophobic layer on both top and bottom substrates with liquid droplet in between them, leads to formation of several capacitors during transport. The capacitance of these capacitors varies as droplet position $x$ changes. 
\end{comment}
Fig. \ref{fig2} shows electrical equivalent circuit for the estimation of voltage drop across the capacitors. The first dielectric and metal layer in the bottom substrate, form two fixed value capacitors ($C_{1c}=C_{2c}=C_{c}$), in addition to three main variable capacitors $C_1(x)$, $C_2(x)$ and $C_3(x)$, found in traditional EWOD system during transport (Fig. \ref{fig3}). The presence of $C_{c}$ in the network, provides direct electrical output terminal $V_0(x)$ proportional to the droplet position $x$ during transition. The voltage drop $V_1(x)$, $V_2(x)$ and $V_3(x)$ are across $C_1(x)$, $C_2(x)$ and $C_3(x)$ respectively. Each capacitor consists of two capacitors in series, one is due to hydrophobic layer ($C_a$) while other due to dielectric layer ($C_b$). \textit{Sensing electrode} and \textit{sensing layer} terms have been used in further discussion to represent control electrode upon which droplet initially rests and insulation layer responsible for $C_{1c}=C_{2c}=C_{c}$ respectively. The subscript $\textsl{1}$ and $\textsl{2}$ represent actuated and unactuated electrode region in the bottom substrate while $\textsl{3}$ represents top grounded electrode region. Similarly, subscript $\textsl{a}$, $\textsl{b}$ and $\textsl{c}$ represent hydrophobic, dielectric and sensing layers respectively.  
\section{Mathematical Modelling}
\label{three}
An analytical energy-based model has been used to evaluate EWOD actuation force on the droplet \cite{Bahadur}. The droplet is assumed to be perfectly conducting and circular in shape during movement. So, there is no voltage drop across the droplet. The total droplet energy $E$ at no actuation condition is given in Eq. \ref{eqn1}: 
\small
\begin{align} 
E&=\gamma_{SL}^{0}\times A_1+\gamma_{SL}^{0}\times A_2\nonumber\\
&\quad+\gamma_{SL}^{0}\times A_3+\gamma_{LA}^{0}\times\ A_{side}
\label{eqn1}
\end{align}
\normalsize
Where, $\gamma_{SL}^{0}$: solid-liquid interfacial energy at 0$V$, $\gamma_{LA}^{0}$: liquid-air interfacial energy at 0$V$, $A_1$: overlapping area between droplet and bottom electrode on which it has to move, $A_2$: overlapping area between droplet and bottom electrode on which it is sitting, $A_3$: overlapping area between droplet and top electrode and $A_{side}$: cylindrical side around the droplet. The following parameters $\gamma_{SL}^{0}$, $\gamma_{LA}^{0}$, $A_1$, $A_2$, $A_3$ and  $A_{side}$ remain constant during no actuation condition and hence the total energy of the droplet remains unchanged and no force acts on the droplet. So, no visible droplet motion is observed.  
\par
When a voltage, $V$ is applied to the right control electrode, the voltage drop across the hydrophobic, insulation and sensing layers causes the reduction of the $\gamma_{SL}^{0}$ according to Lippmann equation (Eq. \ref{eqn2}), while the $\gamma_{LA}^{0}$ and $\gamma_{SA}^{0}$ (solid-air interface energy at 0$V$) remain unchanged.
\small
\begin{align} 
\gamma_{SL}^{V}&=\gamma_{SL}^{0}-\frac{CV^2}{2}
\label{eqn2}
\end{align}
\normalsize
Here $C$ and $V$ are capacitance per unit area $F/m^2$ and potential difference ($p.d.$) across sandwiched dielectric layers respectively while $\gamma_{SL}^{V}$ is solid-liquid interfacial energy at $V$ potential.
\par
The decrease in $\gamma_{SL}^{0}$ as predicted by Lippmann equation induces droplet motion toward the actuating electrode. This leads to a change in $A_1$ and $A_2$ with change in $x$ while $A_3$ remains fixed to value $\pi r^2$ ($r$ is droplet radius). The overlapping areas $A_1(x)$ and $A_2(x)$ for a droplet maintaining a circular shape during transition are given in Eq. \ref{eqn3} \cite{Bahadur}:
\small
\begin{align}
A_1(x)&=r^2\cos^{-1}\left(1-x/r\right)+\left(x-r\right)\sqrt{r^2-\left(x-r\right)^2}\nonumber\\
A_2(x)&=\pi r^2-\alpha_1(x);~~A_3=\pi r^2
\label{eqn3}
\end{align}
\normalsize
\par
The change in $A_1(x)$ and $A_2(x)$ with change in $x$ from $0$ to $L$ (electrode length = $2r$) leads to change in capacitances ($C_1(x)$ and $C_2(x)$, Fig. 2) formed due to hydrophobic and insulation layers. The values of $C_{1}(x)$ and $C_{2}(x)$ with change in $x$ can be expressed as provided in Eq. \ref{eqn4}:
\small
\begin{align} 
\label{eqn4}
\frac{1}{C_1(x)}&=\frac{1}{C_{1a}(x)}+\frac{1}{C_{1b}(x)}=\left\{\frac{s_1+s_2}{\epsilon_0A_1(x)}\right\}\nonumber\\
\frac{1}{C_2(x)}&=\frac{1}{C_{2a}(x)}+\frac{1}{C_{2b}(x)}=\left\{\frac{s_1+s_2}{\epsilon_0A_2(x)}\right\}\\
\frac{1}{C_3}&=\frac{1}{C_{3a}}+\frac{1}{C_{3b}}=\left\{\frac{s_1+s_2}{\epsilon_0A_3}\right\}\nonumber\\
C_{1c}&=C_{2c}=C_{c}=\left\{\frac{\epsilon_0L^2}{s_3}\right\}\nonumber\\
s_1&=t_a/k_a;~~s_2=t_b/k_b;~~s_3=t_c/k_c\nonumber
\end{align}
\normalsize
Where, $\epsilon_{0}$ is vacuum permittivity ($8.854\times10^{-12} F/m$) while, ($t_a$, $k_a$), ($t_b$, $k_b$) and ($t_c$, $k_c$) are thickness and relative permittivity of hydrophobic, insulation and sensing layers respectively. $C_3$ capacitor is formed between droplet and top ground electrode while $C_c$ is the extra capacitor formed due to sensing layer (Fig. \ref {fig1}). These capacitors have been considered to be parallel plate type without taking into account of the fringing effect.
\par
The potential differences ($p.d.$) across these capacitors $V_{1s}(x)$, $V_{2s}(x)$ and $V_{3s}(x)$ are given in Eq. \ref{eqn5}. The numerical value of $V_0(x)$ gives the information about droplet position. Here subscript $s$ has been used to represent $sensing$ scheme.
\small
\begin{align}
\label{eqn5}
V_{1s}(x)&=V\frac{C_2(x)C_3+C_2(x)C_{c}+C_3C_{c}}{\psi(x)}\nonumber\\ 
V_{2s}(x)&=V\frac{C_1(x)C_{c}}{\psi(x)}\\
V_{3s}(x)&=V\frac{C_1(x)\left\{C_2(x)+C_{c}\right\}}{\psi(x)}\nonumber\\
V_0(x)&=V\frac{C_1(x)C_2(x)}{\psi(x)}\nonumber\\
\text{Where,}~\psi(x)&=C_1(x)C_2(x)+C_3\left\{C_2(x)+2C_{c}\right\}\nonumber
\end{align}
\normalsize
\par
Therefore, on applying voltage pulse $V$ to adjacent electrode for droplet transport, the total droplet energy $E(x)$ in the proposed EWOD scheme is given in Eq. \ref{eqn6}:
\small
\begin{align}
\label{eqn6}
E_s(x)&=\left\{\gamma_{SL}^{0}-\frac{C_1(x)V_{1s}(x)^2}{2A_1(x)}\right\}A_1(x)\\
&\quad+\left\{\gamma_{SL}^{0}-\frac{C_2(x)C_{c}\left\{V_{2s}(x)+V_0(x)\right\}^2}{C_2(x)L^2+C_{c}A_2(x)}\right\}A_2(x)\nonumber\\
&\quad +\left\{\gamma_{SL}^{0}-\frac{{C_3}V_{3s}(x)^2}{2A_3}\right\}A_3+\gamma_{LA}^{0}\times A_{side}\nonumber\\
&=E_{1s}(x)+E_{2s}(x)+E_{3s}(x)\nonumber\
\end{align}
\normalsize
Here, $A_{side}$ (cylindrical side around the droplet) has been assumed to remain constant during the transition.
\par
The negative derivative of the total droplet energy $E(x)$ with respect to droplet position $x$, gives the electrostatic actuation force $F_s(x)$ as given in Eq. \ref{eqn7}:
\small
\begin{align}
F_s(x)&=-\frac{dE_s(x)}{dx}=-\left\{\frac{dE_{1s}(x)}{dx}+\frac{dE_{2s}(x)}{dx}+\frac{dE_{3s}(x)}{dx}\right\}\nonumber\\
&=F_{1s}(x)+F_{2s}(x)+F_{3s}(x) 
\label{eqn7}
\end{align}
\normalsize
Reduction of solid-liquid interfacial energy on application of actuating voltage $V$ leads to actuation force $F_s(x)$ acting on droplet during transport. Resultant force $F_s(x)$ consist of three components: $F_{1s}(x)$, $F_{2s}(x)$ and $F_{3s}(x)$; each originates due to change in both capacitance and voltage of $\left\{C_1(x), V_{1s}(x)\right\}$, $\left\{C_2(x), V_{2s}(x)\right\}$ and $\left\{C_3(x), V_{3s}(x)\right\}$ respectively. 
\par
The actuation force $F_w(x)$ in traditional EWOD system is calculated (Eq. \ref{eqn8}) by modifying Eq. \ref{eqn5} and \ref{eqn6}. Here. subscript $w$ has been used to represent traditional EWOD system.
\small
\begin{align}
\label{eqn8}
V_{1w}(x)&=V\frac{C_2(x)+C_3}{C_1(x)+C_2(x)+C_3}\nonumber\\ 
V_{2w}(x)&=V_{3w}(x)=V\frac{C_1(x)}{C_1(x)+C_2(x)+C_3}\nonumber\\
E_w(x)&=\left\{\gamma_{SL}^{0}-\frac{C_1(x)V_{1w}(x)^2}{2}\right\}A_1(x)\\
&\quad +\left\{\gamma_{SL}^{0}-\frac{C_2(x)V_{2w}(x)^2}{2}\right\}A_2(x)\nonumber\\
&\quad +\left\{\gamma_{SL}^{0}-\frac{{C_3}V_{3w}(x)^2}{2}\right\}A_3+\gamma_{LA}^{0}\times A_{side}\nonumber
\end{align}
\normalsize
\par
For better insight in $V_0(x)$ (Eq. \ref{eqn5}, which gives droplet position information in the proposed scheme), values of $C_1(x)$, $C_2(x)$ from Eq. \ref{eqn4} been substituted in Eq. \ref{eqn5} to get the simplified expression of Eq. \ref{eqn9}. 
\small
\begin{align}
\label{eqn9}
V_0(x)&=\frac{V}{2+\frac{A_2(x)}{A_1(x)}+\sigma(x)}\\
\sigma(x)&=\frac{8\pi r^4}{A_1(x)A_2(x)}\beta;~~~\beta=\frac{(s_1+s_2)}{s_3}\nonumber
\end{align}
\normalsize
%%%%%%%%%%%%%%%%%%%%%%%%%%%%%%%%%%%%%%%%%%%%%%%%%%%%%%%%%%%%%%%%%%%%%%%%%%%%%%%
\section{Result and Discussion}
\label{four}
Actuation force $F(x)$ and droplet position sensing output voltage $V_0(x)$ have been simulated during droplet transport by using Eq. \ref{eqn7} and \ref{eqn9} respectively. To get high actuation force, $(s_1+s_2)$ should be very low (see Eq. \ref{eqn4}-\ref{eqn7}); while for linear relation between $V_0(x)$ and droplet position $x$, $s_3$ should be high (Eq. \ref{eqn9}-\ref{eqn10}). So, values of ($t_a$, $k_a$), ($t_b$, $k_b$) and ($t_c$, $k_c$) have been chosen accordingly. Table \ref{table1} outlines the values of system parameters used in simulation. These parameter values are typical of those found in literature \cite{Jean}.
\begin{table}[!h]
\caption{Parameters used for actuation force simulation} % title of Table
\centering % used for centering table
\begin{tabular}{|l| c| c|} % centered columns (4 columns)
\hline %inserts double horizontal lines
Parameter & Value & Unit\\ [0.5ex] % inserts table
%heading
\hline % inserts single horizontal line
Droplet radius $r$ & 1 & $mm$ \\ % inserting body of the table
\hline
Pitch (electrode) length $L$ & 2 & $mm$\\
\hline
Dielectric constant of hydrophobic & & \\
layer $k_a$ ($Teflon~AF~1600$)& 2.0 & -\\ % inserting body of the table
\hline
Dielectric constant & & \\
of insulation layer $k_b$  & 180 & -\\
 ($Barium~Strontium~Titanate$) &  & \\
\hline
Dielectric constant of insulation & & \\
layer $k_c$ ($SiO_2$) & 3.8 & - \\
\hline
Thickness of hydrophobic layer $t_a $ & 20 & $nm$ \\
\hline
Thickness of insulation layer $t_b $ & 70 & $nm$ \\
\hline
Thickness of insulation layer $t_c $ & 1 & $\mu m$ \\ 
\hline
Actuation voltage ($V$) & 50 & $V$\\[1ex] % [1ex] adds vertical space 837 
\hline %inserts single line
\end{tabular}
\label{table1} % is used to refer this table in the text
\end{table}
\begin{figure}[!h]
\begin{center}       
\includegraphics[width=0.9\columnwidth, height=1.950in]{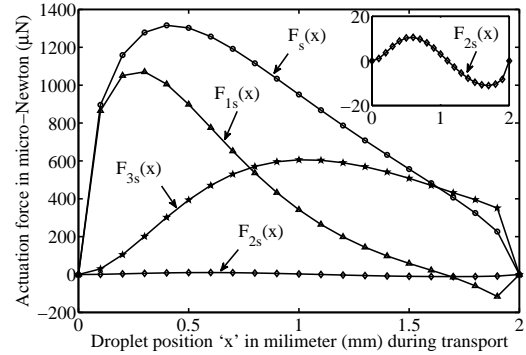}        
\end{center}
\caption{Droplet actuation force and its component during transition in the proposed scheme for droplet position sensing.}
\label{fig4}
\end{figure}
\par
Fig. \ref{fig4} shows profile of resultant actuation force $F_s(x)$ and its components $F_{1s}(x)$, $F_{2s}(x)$ and $F_{3s}(x)$ in the proposed scheme for facilitating position sensing. It can be seen that the resultant force $F_s(x)$ is always positive throughout the range of $x$, which ensures complete transport of the droplet to the actuated electrode. Force components $F_{1s}(x)$ and $F_{3s}(x)$ contribute primarily in $F_s(x)$. $F_{1s}(x)$ provide positive force in the direction of motion for $x\leq 0.825 L$, while $F_{3s}(x)$ provide positive force throughout the range of $x$. Contribution of $F_{2s}(x)$ is negligible. It is important to note that unlike $F_{1s}(x)$ and $F_{3s}(x)$, $F_{2s}(x)$ turns to negative for $x > 0.55 L$ shown in small window in Fig. \ref{fig4}. But because of its very small magnitude compared to $F_{1s}(x)$ and $F_{3s}(x)$, the effect is insignificant.
\begin{figure}[!h]
\begin{center}       
\includegraphics[width=0.9\columnwidth, height=1.95in]{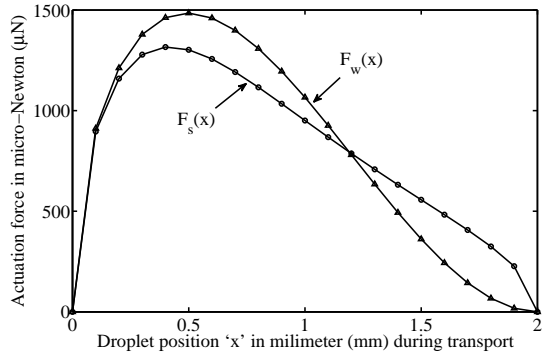}
\end{center}
\caption{Comparison of droplet actuation force in the proposed droplet position sensing scheme ($F_s$) and traditional set-up ($F_w$).}
\label{fig5}
\end{figure}
\par
\begin{figure}[!h]
\begin{center}
\includegraphics[width=0.9\columnwidth, height=1.95in]{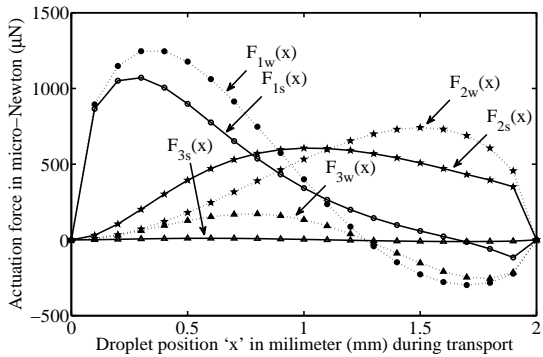}
\end{center}
\caption{(a): Comparison of component force profiles ($F_1(x)$, $F_2(x)$ and $F_3(x)$) in the proposed droplet position sensing scheme and traditional set-up.}
\label{fig6}
\end{figure}

Actuation force in traditional EWOD scheme has also been simulated by using Eq. \ref{eqn7}-\ref{eqn8} with same parameters value (Table \ref{table1}), in order to compare its force profile with the proposed scheme. The force profile for $F(x)$ in both the schemes are found to be similar. Moreover, profile of $F_1(x)$, $F_2(x)$ and $F_3(x)$ are also similar in both the cases as shown in Fig \ref{fig6}.  
\begin{figure}[!h]
\begin{center}    
\includegraphics[width=0.9\columnwidth, height=1.95in]{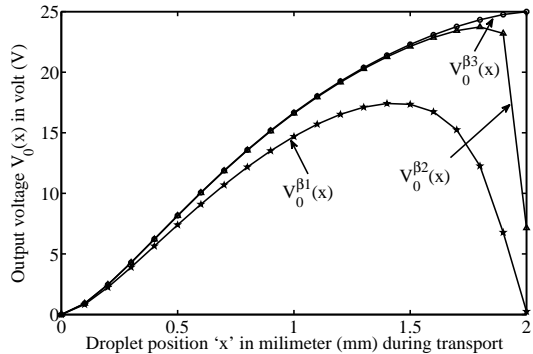}       
\end{center}
\caption{Variation of output voltage $V_0(x)$ with droplet position $x$ during transport in the proposed scheme. Here $V_0^{\beta1}(x)$, $V_0^{\beta2}(x)$ and $V_0^{\beta3}(x)$ represents $V_0(x)$ profile when $\beta_1=39.5\times10^{-3}$, $\beta_2=1\times10^{-3}$ and $\beta_3=1\times10^{-6}$ respectively.}
\label{fig7}
\end{figure}
\par
Analysis of voltage variation of $V_0(x)$ (with droplet position $x$), during transition has been carried out by simulating Eq. 10 with same value of the parameters (outlined in Table 1). Fig. \ref{fig7} ($V_0^{\beta_1}(x)$) shows the output voltage $V_0(x)$ with droplet position $x$. It can be seen that output voltage $V_0(x)$ is almost linearly proportional with droplet position $x$ upto 1.4 $mm$ and can be directly used for position sensing in this range (i.e $x\leq 1.4$ $mm$) without using any further signal conditioning circuit. The observed nonlinearity in $V_0^{\beta_1}(x)$, is due to low value of $\beta$ ($39.5\times10^{-3}$) based on chosen parameter value (Table \ref{table1}).
\par
In Eq. \ref{eqn9}, $\sigma(x)$ is dependent on system parameters, and also responsible for the observed nonlinearity found in $V_0(x)$ as shown in Fig. \ref{fig7}. This nonlinearity can be removed by making $\sigma(x)$ negligibly small. Hence, in order to linearize $V_0(x)$, Eq. \ref{eqn9} can be rewritten as:
\small
\begin{align}
V_0(x)&=\frac{V}{2+\frac{A_2(x)}{A_1(x)}}
\label{eqn10}
\end{align}
\normalsize
Where, $\sigma(x)$ has been assumed to be zero. Plot of $V_0(x)$ by simulating Eq. \ref{eqn10} has been shown in Fig. \ref{fig7} ($V_0^{\beta_3}(x)$). The curve of $V_0(x)$ is almost linear throughout the range of $x$ ($0-L$). In addition, $V_0(x)$ in Eq. \ref{eqn10} has become independent of system parameters $t_a$, $k_a$, $t_b$, $k_b$, $t_c$ and $k_c$, making the sensing scheme insensitive to variation in these parameters with change in ambient condition and aging. The dependency of $V_0(x)$ on amplitude of the voltage pulse $V$ and radius of droplet $r$ modifies only sensitivity and slope of profile respectively, but linearity remains unaffected.
\par
$\sigma(x)$ term can not be made absolutely equal to zero throughout the range of $x$ ($0-L$) by adjusting non dimensional parameter $\beta$ given in Eq. \ref{eqn9}, but can be reduced to a value, so that an almost linear response can be obtained. It has been found that for $\beta \leq 1\times10^{-6}$, the effect of $\sigma(x)$ becomes insignificant in Eq. \ref{eqn9} and a linear response is obtained from Eq. \ref{eqn10} (Fig. \ref{fig7}, $V_0^{\beta_3}(x)$). Adjusting system parameters $t_a$, $k_a$, $t_b$, $k_b$, $t_c$ and $k_c$ in order to get a $\beta$ value of the order of $10^{-6}$ may be difficult from implementation point of view. But a linear $V_0(x)$ response with droplet position $x$ upto $0.9L$ can easily be found, if $\beta \leq 1 \times10^{-3}$ (Fig. \ref{fig7}, $V_0^{\beta_2}(x)$). A $39.5~\mu m$ ($t_c$) thick film of sensing layer ($SiO_2$) is required to achieve a $\beta$ value of $1\times10^{-3}$.
\section{Conclusion}
\label{five}
A scheme for monitoring multiple droplets position simultaneously in electrowetting-on-dielectric (EWOD) based devices has been proposed. Mathematical modelling of the proposed scheme for EWOD force $F(x)$ and position sensing output voltage $V_0(x)$ has been carried out based on energy minimization approach. The actuation force profile in this scheme is positive as well as comparable to those found in traditional set-up throughout the transition length ensuring complete droplet transfer.  A direct voltage output having almost linear relation with droplet position $x$ is found with proper system parameters value. It provides better isolation between actuation and sensing line as well.
\section*{Acknowledgment}
The authors are thankful to Department of Biotechnology, Government of India for financial assistance (Project: BT/PR3713/MED/32/203/2011) to carry out the work.
\bibliographystyle{IEEEtr}
\bibliography{sensor}

\begin{thebibliography}{1}

\bibitem{Jean}
J.~Berthier, {\em Microdrops and Digital Microfluidics}.
\newblock William Andrew Inc., 2008.

\bibitem{Shin}
Y.~J. Shin and J.~Lee, ``Machine vision for digital microfluidics,'' {\em
  Review of Scientific Instruments}, vol.~81, pp.~014302 (1--8), 2010.

\bibitem{Shih}
Y.~J. Shin and J.~Lee, ``A feedback control system for high-fidelity digital
  microfluidic,'' {\em Lab Chip}, vol.~11, pp.~535--540, 2011.

\bibitem{Schertzer}
M.~J. Schertzer, R.~Ben-Mrad, and P.~E. Sullivan, ``Automated detection of
  particle concentration and chemical reactions in ewod devices,'' {\em Sensors
  and Actuators B: Chemical}, vol.~164, pp.~1--6, 2012.

\bibitem{Murran}
M.~A. Murran and H.~Najjaran, ``Capacitance-based droplet position estimator
  for digital microfluidic devices,'' {\em Lab Chip}, vol.~12, pp.~2053--2059,
  2012.

\bibitem{Bhattacharjee}
B.~Bhattacharjee and H.~Najjaran, ``Droplet sensing by measuring the
  capacitance between coplanar electrodes in a digital microfluidic system,''
  {\em Lab Chip}, vol.~12, pp.~4416--4423, 2012.

\bibitem{Sadeghi}
S.~Sadeghi, H.~Ding, S.~C. G.~J.~Shah, P.~Y. Keng, C.~Kim, and R.~M.~V. Dam,
  ``On chip droplet characterization: a practical, high-sensitivity measurement
  of droplet impedance in digital microfluidics,'' {\em Analytical Chemistry},
  vol.~84, 2012.

\bibitem{Luan2012}
L.~Luan, M.~W. Royal, R.~Evans, and R.~B. Fair, ``Chip scale optical
  microresonator sensors integrated with embedded thin film photodetectors on
  electrowetting digital microfluidics platforms,'' {\em IEEE Sensor Journal},
  vol.~12, 2012.

\bibitem{Bahadur}
V.~Bahadur and S.~V. Garimella, ``An energy-based model for
  electrowetting-induced droplet actuation,'' {\em J. Micromech. Microeng.},
  vol.~16, pp.~1494--1503, 2006.

\end{thebibliography}
\end{document}